\documentclass{article}
\usepackage{aaai20}
\usepackage{times}  
\usepackage{graphicx}
\usepackage{color}
\usepackage{amsthm}
\usepackage{url}

\title{Avoiding Negative Side Effects due to Incomplete Knowledge of AI Systems}
\nocopyright

\author{Sandhya Saisubramanian\textsuperscript{1} \and Shlomo Zilberstein\textsuperscript{1} \and Ece Kamar\textsuperscript{2} \\[2pt]
	\textsuperscript{1}{College of Information and Computer Sciences, University of Massachusetts Amherst, Massachusetts, USA}\\
	\textsuperscript{2}{Microsoft Research, Redmond, Washington, USA}}
\begin{document}
	
	\maketitle

\begin{abstract}
Autonomous agents acting in the real-world often operate based on models that ignore certain aspects of the environment. The incompleteness of any given model---handcrafted or machine acquired---is inevitable due to practical limitations of any modeling technique for complex real-world settings. Due to the limited fidelity of its model, an agent's actions may have unexpected, undesirable consequences during execution. Learning to recognize and avoid such \emph{negative side effects} of an agent's actions is critical to improve the safety and reliability of autonomous systems. Mitigating negative side effects is an emerging research topic that is attracting increased attention due to the rapid growth in the deployment of AI systems and their broad societal impacts. This article provides a comprehensive overview of different forms of negative side effects and the recent research efforts to address them. We identify key characteristics of negative side effects, highlight the challenges in avoiding negative side effects, and discuss recently developed approaches, contrasting their benefits and limitations. The article concludes with a discussion of open questions and suggestions for future research directions.

\end{abstract}

\section{Introduction}
A world populated with intelligent and autonomous systems that simplify our lives is gradually becoming a reality. These systems are \emph{autonomous} in the sense that they can devise a sequence of actions to achieve some given objectives or goals, without human intervention. Such systems are deeply integrated into our daily lives through various applications such as mobile health monitoring~\cite{sim2019mobile}, intelligent tutoring~\cite{folsom2013tractable}, self-driving cars~\cite{Zaaai15}, and clinical decision making~\cite{bennett2013artificial}. This broad deployment brings along new challenges and increased responsibility for designers of AI systems, particularly ensuring that these systems operate as expected when deployed in the real-world. Despite recent advances in artificial intelligence and machine learning, there are no ways to assure that systems will always ``do the right thing'' when operating in the open world~\cite{lakkaraju2017identifying}.

For example, consider an autonomous vehicle (AV) that was carefully designed and tested for safety aspects such as yielding to pedestrians and conforming to traffic rules. When deployed, the AV may not slow down when driving through puddles and splash water on nearby pedestrians. Another documented example of undesirable behavior in AVs is the vehicle swerving left and right multiple times to localize itself for active lane-keeping. During this process, the vehicle rarely prompted the driver to take control~\cite{iihsreport}. This behavior, especially on curvy and hilly roads, can startle the driver or cause panic.

Undesirable behaviors may occur even when performing relatively simple tasks. For example, robot vacuum cleaners are becoming increasingly popular and they have a simple task---to remove dirt from the floor. A robot vacuum cleaner in Florida ran over animal feces in the house and continued its cleaning cycle, smearing the mess around the house~\cite{roombapoop}. In an extreme case in South Korea, a robot vacuum cleaner locked into the hair of a woman who was sleeping on the floor, mistaking her hair for dust~\cite{vacuumtragedy}. 

A key factor affecting an agent's performance is its knowledge of the environment in which it is situated. In these examples, the agent was performing its task, perhaps optimally with respect to the information provided to it, but there were serious negative side effects to the agent's actions. In the AV example, driving fast through puddles is optimal when optimizing travel time. The side effects are due to the limited scope of the agent's model, not accounting for the undesirability of splashing water on pedestrians. In practice, it is not feasible to anticipate all possible negative side effects and accurately encode them in the model at design time. Due to the practical limitations of data collection and model specification, agents operating in the open world often rely on incomplete knowledge of their target environment which may lead to unexpected, undesirable consequences. Addressing the potential undesirable behaviors of autonomous systems is critical to support long-term autonomy and ensure that a deployed AI system is reliable. 

There have been numerous recent studies focused on the broad challenge of building safe and reliable AI systems~\cite{amodei2016concrete,russell2015research,saria2019tutorial,thomas2019preventing}.  Here, we examine the particular problem of identifying and mitigating the impacts of undesirable side effects of an agent's actions when operating in the open world.  We do not consider system failure or negative side effects that result from intentional adversarial attack on the system~\cite{biggio2018wild,cao2019adversarial}.

\begin{figure}
	\centering
	\includegraphics{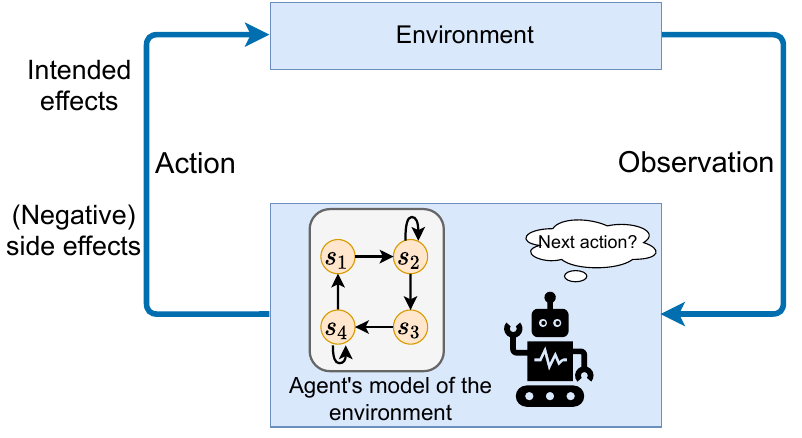}
	\caption{Negative side effects of an agent's behavior.}
	\label{fig:illus}
\end{figure}

\begin{quote}
	\emph{\textbf{Negative side effects} (NSE) are undesired effects of an agent's actions that occur in addition to the agent's intended effects when operating in the open world.} (Figure~\ref{fig:illus}).
\end{quote}

Negative side effects occur because the agent's model and objective function focus on some aspects of the environment but its operation could impact additional aspects of the environment. The value alignment problem studies the unsafe behavior of an agent when its objective does not align with human values~\cite{hadfield2016cooperative,russell2017provably,russell2019human}. Misaligned systems are more likely to produce negative side effects. However, the occurrence of negative side effects does not necessarily indicate that there is a value alignment problem.  Negative side effects can occur even in settings where the agent optimizes legitimate objectives that align with the user's goals, due to incomplete knowledge and distributional shift. For example, while driving in Boston, AVs that are programmed to not run into obstacles were stopped by the local breed of unflappable seagulls standing on the street~\cite{nutonomy}. Not running into obstacles is well-aligned with the users' intentions and objectives, but there are side effects because the agent lacks knowledge that it can edge to startle the birds and then continue driving. In fact, such knowledge was later added to the system to resolve the problem. In addition, some systems may cause unavoidable negative side effects that cannot be mitigated. While the side effects may be undesirable, the user may accept the system as is, once they learn about it and recognize that the side effects are unavoidable. In such cases, we cannot say that there is a value alignment problem, even though the negative side effects may occur.

Certainly, some negative side effects could be anticipated or detected during system development and appropriate mechanisms to mitigate their impacts could be implemented prior to deployment.  This article focuses on negative side effects that are discovered when the system is deployed, due to a variety of factors such as unanticipated domain characteristics, unanticipated consequences of system or software upgrade, or cultural differences among the target user and development team. Design decisions that may be innocuous during initial testing may have a significant impact when a system is widely deployed. For example, the issue of a Roomba locking into the hair of a person lying on the floor emerged only after the system was deployed in Asia. Overcoming negative side effects is an emerging area that is attracting increased attention within the AI community~\cite{amodei2016concrete,hadfield2017inverse,hibbard2012avoiding,krakovna2019penalizing,russell2017provably,turner2020conservative,SKZijcai20,shah2019preferences,zhang2018minimax}.

The severity of negative side effects may range from mild to safety-critical failures.
Often, the discussions around the risk of encountering negative side effects have highlighted catastrophic events.  While these discussions are critical and essential, AI systems in general are carefully designed and tested for such failures before deployment. With the increasing growth in the capabilities and deployment of AI systems, it is equally important to address the negative side effects that are not catastrophic, but have significant impacts. Such side effects occur more frequently but are often overlooked, particularly when the only remedy available is to remove the product and develop a new version that can avoid the undesired behavior.  Hence, providing end users the tools to identify and mitigate the impacts of negative side effects 
is critical in shaping how users view, interact, collaborate, and trust AI systems~\cite{saisubramanian2021understanding}.  

The rest of this article identifies key characteristics of negative side effects, highlights the challenges in overcoming negative side effects, and discusses the recent research progress in this area. To promote a better understanding of the prevalence of negative side effects and to provide common test cases for the research community, we have created a public repository that allows AI researchers to report new cases.  We conclude the article with a discussion of open questions to encourage future research in this area.

We introduce a taxonomy of negative side effects, outlined in Table~\ref{tab:classification}. Understanding the characteristics of negative side effects (NSE) helps design better solution approaches to detect and mitigate their impacts in deployed systems.

\begin{table}
	\centering 
	\renewcommand{\arraystretch}{1.5}
	\begin{tabular}{|l|p{5cm}|}
		\hline
		Property  &  Property Values 
		\\ \hline \hline
		Severity & Ranges from mild to safety-critical  \\ \hline 
		
		Reversibility & Reversible or irreversible	\\ \hline 
		
		Avoidability & Avoidable or unavoidable \\ \hline
		
		Frequency & Common or rare \\ \hline
		
		Stochasticity & Deterministic or probabilistic  \\ \hline
		
		Observability & Full, partial, or unobserved \\ \hline
		
		Exclusivity & Prevent task completion or not  \\  
		\hline
	\end{tabular}
	\caption{Taxonomy of negative side effects.}
	\label{tab:classification}
\end{table}

\vspace{4pt}
\noindent \textbf{Severity:} The severity of negative side effects ranges from mild side effects that can be largely ignored to safety-critical failures that require suspension of the system deployment. Safety-critical side effects are typically addressed by redesigning the model and hence require extensive evaluation before redeployment.  An example of a safety-critical NSE is an AV failing to detect a construction worker's hand gestures~\cite{crane2016survey}.
We conjecture that many negative side effects lie in the middle with significant impacts that require attention, but not sufficiently critical to suspend the service.  An autonomous vehicle that does not slow down when going through puddles can cause significant impacts, but those are unlikely to be considered sufficiently critical to roll-back its deployment, particularly if mechanisms are provided to mitigate the negative impacts.  Addressing such NSE without suspension of service requires agent adaptation and online planning.

\vspace{4pt}	
\noindent \textbf{Reversibility:} Side effects are reversible if the impact can be reversed or negated, either by the agent causing it or via external intervention. For example, breaking a vase is an irreversible side effect, regardless of the agent's skills~\cite{amodei2016concrete}. Side effects such as leaving marks on a wall can be fixed by repainting it, but the agent may require external assistance to achieve that. 

\vspace{4pt}
\noindent \textbf{Avoidability:} In some problems, it may be impossible to avoid the negative side effects during the course of the agent's operation to complete its assigned task. This introduces a trade-off between performing agent's assigned task and avoiding the side effects. For example, the side effects of driving through puddles are unavoidable if all roads leading to the destination have puddles. Addressing unavoidable NSE requires a principled approach to balance the trade-off between avoiding side effects and optimizing the completion of the assigned task. 

\vspace{4pt}	
\noindent \textbf{Frequency:} The frequency of occurrence of negative side effects depends on the environmental conditions and the action plan. Certain NSE may occur rarely, considering all use cases, but may occur frequently for a small subset of cases. A robot pushing a box over a rug may dirty it as a negative side effect. This is an example of a frequently occurring negative side effect when the domain of operation is largely covered with a rug. 
The frequency of occurrence could impact the approach to identify negative side effects and the corresponding mitigation approach.

\vspace{4pt}	
\noindent \textbf{Stochasticity:} The occurrence of negative side effects may be deterministic or probabilistic. Deterministic NSE always occur when some action preconditions arise in the open world. Side effects are probabilistic when their occurrence is not certain even when the right preconditions arise. For example, there may be a small probability that a robot may accidentally slide and scratch the wall while pushing a box, but that undesired effect may happen only 20\% of the times the robot slips.  

\vspace{4pt}	
\noindent \textbf{Observability:} The agent's observability of the actual NSE or the conditions that trigger them are generally determined by the agent's state representation and sensory input.  The side effects may be fully observable, partially observable, or even unobserved by the agent. Observing a side effect is different from identifying or recognizing the impact as a side effect. For example, the agent may observe the scratch it made on the wall but may not be aware that it is undesirable, and as a result may not try to avoid it. Observability is a critical factor when learning to avoid NSE.  When an external authority provides feedback to the agent, it may be sufficient for the agent to observe the conditions that trigger the negative side effect.  However, when an agent may need to identify NSE on its own, it needs more complex general knowledge about the open world.

\vspace{4pt}	
\noindent \textbf{Exclusivity:} Negative side effects may prevent the agent from completing its assigned task. This category is relatively easier to identify. Often, however, the side effects negatively impact the environment without preventing the agent from completing its assigned task. Such side effects are typically difficult to identify at design time. Much of the current research on avoiding negative side effects focuses on side effects that do not prevent the agent from completing its current primary task.

\section{Challenges in Avoiding Negative Side Effects}
The challenges in avoiding negative side effects broadly stem from the difficulty in obtaining knowledge about NSE a priori, gathering user preferences to understand their tolerance for side effects, and balancing the potential trade-off between completing the task and avoiding the side effects.

\paragraph{Model imprecision}  Agents designed to operate in the open world are either trained in a simulator, or operate based on models created by a designer or generated automatically using data. Regardless of how much effort goes into the system design and how much data is available for training and testing, it is generally infeasible to obtain a perfect description of open-world environments. Practical challenges in model specification, such as the qualification and ramification problems, and computational complexity consideration often cause the agent to reason based on models that do not represent all the relevant details in the open world~\cite{dietterich2017steps}. Simulators also suffer from this drawback, as they are also built by designers, resulting in mismatches between a simulator and the actual environment~\cite{ramakrishnan2019overcoming}. As a result of reasoning with incomplete information, agents may not consistently behave as intended, leading to unexpected and costly errors, or may completely fail in complex settings.

There are three key reasons why the agent may not have prior knowledge about the negative side effects of its actions. First, identifying NSE \emph{a priori} is inherently challenging. As a result, this information is often lacking in the agent's model. Second, many AI systems are deployed in a variety of settings, which may be different from the environment used in training and testing of the agent. This \emph{distributional shift} may cause NSE and is difficult to assess during the design process. Third, negative side effects in many settings arise due to \emph{user preference} violation. It is generally difficult to precisely learn or encode human preferences and account for individual or cultural differences.

Techniques such as online model update and policy repair to minimize side effects, and building more realistic simulators~\cite{dosovitskiy2017carla} are some of the promising directions to handle negative side effects due to model imprecision.

\paragraph{Feedback collection}  An agent that is unaware of the side effects of its actions can gather this information 
through feedback from users or through autonomous exploration and model revisions. Though learning from feedback produces good results in many problems~\cite{lakkaraju2017identifying,ramakrishnan2019overcoming,SKZijcai20,zhang2018minimax,zhangquerying,Baamas20}, there are three main challenges in employing this approach in real-world systems. First, the learning process may not be \emph{sample efficient} or may require feedback in a certain format to be sample efficient, such as correcting the agent policy by providing alternate actions for execution. Feedback collection in general is an expensive process, particularly when the feedback format requires constant human oversight or imposes significant cognitive overload on the user. Second, feedback may be \emph{biased} or \emph{delayed} or both, which in turn affects the agent's learning process. Finally, it is generally assumed that the agent uses \emph{human-interpretable representations} for querying and feedback collection, but there may be mismatches between the models of the agent and human. There are some recent efforts towards addressing the problem of sample efficiency in learning~\cite{buckman2018sample,wang2016sample} and investigating the impact of bias in feedback for agent learning~\cite{ramakrishnan2018discovering,SKZijcai20}. Identifying and evaluating human-interpretable state-action representations for querying humans is largely an open problem.

\paragraph{Managing tradeoffs} When negative side effects are unavoidable and interfere with the performance of the agent's assigned task, there is a trade-off between completing the task efficiently and avoiding the NSE. In an extreme case, it may be impossible for the agent to achieve its goal without creating negative side effects.  How far should an agent deviate from its optimal plan in order to minimize the impacts of negative side effects? Balancing this trade-off requires user feedback since it depends on their tolerance for negative side effects. This can be challenging when the agent's objective and the side effects are measured in different units.

\section{Approaches to Mitigate Negative Side Effects} 
This section reviews the emerging approaches to mitigating the impacts of negative side effects. Table~\ref{tab:summary} summarizes the characteristics of side effects handled by each one of the methods we mention.

\begin{table*}[t]
	\centering 
	\renewcommand{\arraystretch}{1.25}
	\resizebox{\textwidth}{!}{%
		\begin{tabular}{|l|c|c|c|c|c|c|c|}
			\hline
			& Severity  &  Reversibility & Avoidability & Frequency & Stochasticity & Observability & Exclusivity 
			\\ \hline \hline
			
			[Hadfield-Mennel et al., 2017] & - & irreversible & - & frequent & deterministic & - & -  \\
			
			[Zhang et al., 2018] & - & irreversible & avoidable & - & deterministic & observable & non-interfering \\ 
			
			[Krakovna et al.,2019] & - & - & - & - & - & observable & non-interfering \\
			
			[Shah et al., 2019] & - & irreversible & - & frequent & deterministic & observable & non-interfering \\
			
			[Zhang et al., 2020] & - & irreversible & - & - & deterministic & observable & - \\
			
			[Turner et al., 2020a] & - & irreversible & avoidable & frequent & deterministic & - & non-interfering \\
			
			[Saisubramanian et al., 2020] & not safety-critical & irreversible & - & frequent & deterministic & - & non-interfering \\
			
			[Turner et al., 2020b] & - & - & - & frequent & deterministic & - & - \\
			
			[Krakovna et al., 2020] & not safety-critical & - & - & - & - & observable & - \\
			
			[Saisubramanian et al., 2021] & - & - & - & frequent & deterministic & - & non-interfering \\

			\hline
		\end{tabular}
	}
	\caption{Summary of the characteristics of the surveyed approaches to mitigate negative side effects. ``-'' indicates the approach is indifferent to the values of that property. Although some existing works do not explicitly refer to the severity of the side effects they can effectively handle, in general these approaches target side effects that are undesirable and significant, but not safety-critical. 
	}
	\label{tab:summary}
\end{table*}

\paragraph{Model and policy update}  The occurrence of negative side effects in a system depends on the agent's trajectory, which is determined by its policy derived using its reasoning model. Hence, a natural approach to mitigate NSE is to update the model such that the agent's policy avoids NSE as much as possible. When the side effects are safety-critical, the model update may include significant changes such as redesign of the reward function.  \citeauthor{hadfield2017inverse}~(\citeyear{hadfield2017inverse}) address such a setting where the negative side effects occur due to unintentional misspecification of rewards by the designer. It is assumed that the designer prescribes a proxy reward function and the agent is assumed to be \emph{aware} of a possible reward misspecification. 
The proxy reward function is treated as a set of demonstrations, and the agent learns the intended reward function using approximate solutions for inference. 
As acknowledged by the authors, this approach is not scalable to large, complex settings.

Redesigning the reward function may degrade the agent's performance with respect to its assigned task or introduce new risks, and hence requires exhaustive evaluation before redeployment. This could be very expensive and likely require suspension of operation until the newly derived policies could be deemed safe for autonomous operation. In problem domains where the side effects are undesirable but not safety-critical, the impact can be minimized by augmenting the agent's model with a penalty function corresponding to NSE. This exploits the reliability of the existing model with respect to the agent's assigned task, while allowing a deployed agent to adjust its behavior to minimize the side effects. 

In related work~\cite{SKZijcai20}, we describe a multi-objective formulation of this problem with a lexicographic ordering of objectives that prioritizes optimizing the agent's assigned task (primary objective) over minimizing NSE (secondary objective). A slack value to the primary objective determines the maximum allowed deviation from the optimal expected reward of the primary objective so as to minimize side effects. This work considers a setting in which the agent has \emph{no prior knowledge} about the side effects of its actions. Information about NSE is gathered using feedback, which is then encoded by a reward function. The agent may not be able to observe the NSE except for the penalty, which is proportional to the severity of the NSE provided by the feedback mechanism. The model is updated with this learned reward function and an updated policy is computed that avoids negative side effects as much as possible, within the allowed slack. This formulation can hence handle both avoidable and unavoidable NSE. However this approach is not suitable for safety-critical consequences since it prioritizes optimizing the completion of the agent's assigned task.

Both these approaches address the side effects associated with the execution of an action, independent of its outcome.

\paragraph{Constrained optimization}  Negative side effects occur when an agent alters features in the environment that the user does not expect or desire to be changed. This can be addressed by constraining the features that can be altered by the agent during its operation. In~\cite{zhang2018minimax}, the authors consider a setting in which the uncertainty over the desirability of altering a feature is included in the agent's model and considers deterministic side effects that are irreversible, but avoidable. The agent first computes a policy assuming all the uncertain features are ``locked'' for alteration. If a policy exists, then the agent executes it. If no policy exists, the agent queries the human to determine which features can be altered and recomputes a policy. A regret minimization approach is used to select the top-$k$ features for querying. Recently, the authors extended this approach to identify if NSE are unavoidable by casting it as a set-cover problem~\cite{zhangquerying}. If the side effects are unavoidable, the agent ceases operation. Therefore, these approaches are not suitable for settings where the agent is expected to alleviate (unavoidable) NSE to the extent possible, while completing its assigned task.

\paragraph{Minimizing deviations from a baseline}  Another class of solution methods defines a penalty function for negative side effects as a measure of deviation from a baseline state, based on the features altered. The deviation measure reflects the degree of disruption to the environment caused by the agent's actions. The agent is expected to minimize the disruption while pursuing its goal, thereby mitigating NSE. In~\cite{krakovna2019penalizing}, the authors present a multi-objective formulation with scalarization, with the deviation from baseline state measured using reachability-based metrics. The agent's sensitivity to NSE can be adjusted by tuning the scalarization parameters. The relative reachability approach~\cite{krakovna2019penalizing} is not straightforward to apply in settings more complex than grid-worlds, as acknowledged by the authors. Furthermore, the resulting performance is sensitive to the metric used to calculate deviations, particularly the choice of baseline state.

Different candidates for baseline states have been proposed, such as start state and inaction in a state~\cite{krakovna2019penalizing}. These baselines do not consider human preferences and may penalize all side effects. To overcome this,  \citeauthor{shah2019preferences} (\citeyear{shah2019preferences}) present a Maximum Causal Entropy approach to infer human preferences from the start state. They assume that an environment is typically optimized for human preferences and the agent can mitigate NSE by inferring human preferences before it starts acting.  This approach, however, requires knowledge about the dynamics of the environment  to determine if the environment has been optimized for human preferences or not. 

\paragraph{Human-agent collaboration} Approaches such as policy update, constrained optimization, and minimizing deviations from a baseline rely heavily on the fidelity of agent's state representation. In many cases, however, the agent's state representation may only include the features relevant to its assigned task. This limited representation can impact the agent's ability to learn and mitigate NSE. In recent work~\cite{saisubramanian2021mitigating}, we describe a human-agent team approach that mitigates NSE via environment shaping. Environment shaping is the process of applying simple modifications to the current environment to make it more agent-friendly and minimize the occurrence of side effects. The agent optimizes its assigned task, unaware of the side effects of its actions. The human mitigates the side effects of the agent through simple reconfigurations of the environment. This approach is applicable to settings where the user can assist the agent actively, beyond providing feedback, and there are one or more agents with limited state representation. This approach is not suitable for environments that are not configurable by the user or when the agent's model and policy are frequently updated.

\paragraph{Accounting for auxiliary objectives and future tasks}
Attainable utility~\cite{turner2020conservative,turner2020avoiding} measures the impact of side effects as the shifts in the agent's ability to optimize for auxiliary objectives, generalizing the relative reachability measure. 
Often, the occurrence of NSE may not impact the agent's ability to complete its current assigned task, but may affect future task completion. To minimize the interference with future tasks, \citeauthor{krakovna2020avoiding}~(\citeyear{krakovna2020avoiding}) present an approach that provides the agent an auxiliary reward for preserving agent ability to perform future tasks in the environment. These approaches assume that the agent's state representation is sufficient to calculate the deviations and are therefore not directly applicable to settings with mismatches between the agent's state representation and the environment. 

\begin{figure}
	\centering
	\includegraphics{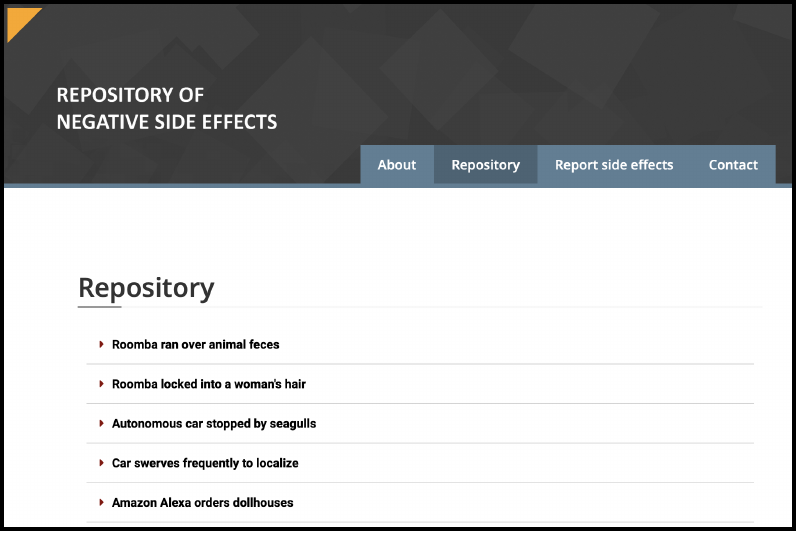}
	\caption{A public repository of negative side effects}
	\label{fig:repo}
\end{figure}

\section{A Repository of Negative Side Effects}
Since the problem of negative side effects is an emerging topic, current research relies on proof-of-concept toy domains for performance evaluation. Moving forward, understanding the occurrence of negative side effects in deployed AI systems is necessary for a realistic formulation of the problem and to design effective solution approaches to address it. To that end, we have created a repository of negative side effects~\cite{nse-repository}. This publicly available repository is shown in Figure~\ref{fig:repo}. It contains real-world instances from scientific reports or news articles, identified by us. For each instance, details such as problem setting in which negative side effects were observed, a description of the side effects, location and date of incident, are provided. We believe this repository will promote a deeper understanding of the problem, provide insights about which assumptions are valid, and facilitate moving beyond simple grid-world type domains as common test cases to evaluate techniques. 

We invite the readers to contribute to this repository by reporting cases of negative side effects of deployed AI systems, based on user experiences, published papers, or media reports, using an online form we provide~\cite{nse-form}. Each submission will be reviewed by our team before adding it to the repository. 
	
\section{Open Questions and Future Work}
Some key open questions and research directions that can further the understanding of negative side effects and development of strategies to mitigate their impacts are discussed below.

\vspace{4pt}
\noindent \textbf{Negative side effects in multi-agent settings: } The existing works have studied the negative side effects of a single agent's actions on the environment. In collaborative multi-agent systems, agents work together to optimize performance and may have complementary skills. For example, the negative side effects produced by an agent may be reversible by another agent. \emph{How can we leverage collaborative multi-agent settings to effectively mitigate negative side effects?} One solution approach is to devise a joint policy to mitigate the negative side effects, in addition to optimizing the utility of the assigned task. The existing rich body of work on cooperative multi-agent systems examines how the intended effects of each agent's actions may affect the other agents when devising a joint policy that maximizes the performance~\cite{pynadath2002multiagent,GZaamas03,zhang2007meta,ramakrishnan2019overcoming}.  Extending such frameworks to handle the side effects problem requires knowledge about the negative side effects of each agent's actions and how it affects the behavior and rewards of other agents in the environment.
External feedback may indicate the occurrence of NSE as a result of a joint action of the agents. Effectively mitigating the side effects requires mechanism design for precise identification of the agent whose actions produce these undesirable effects, based on the feedback provided for joint actions.

\vspace{4pt}
\noindent \textbf{Addressing side effects in partially observable settings: } 
In partially observable settings, an agent operates based on a belief distribution over the states. The problem is further complicated when the agent has no prior knowledge of the side effects, which may be partially observable or unobserved. \emph{How can an agent effectively learn to avoid negative side effects in partially observable settings?} Due to partial observability, the agent maps the external feedback indicating the occurrence of negative side effects to a belief distribution and not an exact state. As a result, a belief distribution may be associated with multiple conflicting feedback. Depending on how the feedback signals are aggregated, different types of agent behavior emerge with varying sensitivity to negative side effects.

\vspace{4pt}
\noindent \textbf{Combination of side effects: } Many AI systems, such as autonomous vehicles, are comprised of multiple entities that function together to achieve a goal. Each of these entities may contribute to different forms of negative side effects. It is likely that multiple forms of negative side effects, with varying impacts and severity, co-exist and require different solution techniques to mitigate the overall impact. \emph{How to ensure that approaches designed to eliminate one form of side effect do not introduce new risks?} This problem is related to avoiding negative side effects in collaborative multi-agent settings since each component can be treated as an agent collaborating with other agents. Reasoning about multiple forms of risks together is a cornerstone in achieving safe AI systems. One approach is to evaluate the effects of an impact regularizer on other modules in the system that interact with the module of interest. This requires broad background knowledge about the architecture and functionality of each component, which may not be available in systems with black-box components.

\vspace{4pt}
\noindent \textbf{Skill discovery to mitigate negative side effects: }
Skill discovery~\cite{eysenbach2018diversity,konidaris2009skill} in reinforcement learning allows an agent to discover useful new skills autonomously. High-level skills or \emph{options} are temporally extended courses of actions that generalize primitive actions of an agent. These closed-loop policies speed up planning and learning in complex environment and are generally used in hierarchical methods for reasoning. Exploring the feasibility of \emph{skill discovery for avoiding negative side effects} is an interesting direction that could accelerate agent behavior adaptation, especially to avoid side effects during agent exploration. For example, if the agent learns to push a box without scratching the walls or dirtying the rug, this option is useful in a variety of related settings and enables faster behavior adaptation.

\vspace{4pt}
\noindent \textbf{Beyond Safety and Control: }
This article has discussed the undesirable side effects in the context of safety and control in embodied autonomous systems. Investigating negative side effects of decision-support systems and recommender systems is an important direction for the future. Negative side effects in these contexts may not be immediate, such as the effect on climate change, human health, or cognitive ability caused by the system's decisions. 

AI systems may also suffer from other factors that affect their reliability, such as biases and privacy concerns. Amplifying underlying biases in a system or increased vulnerability to attacks may occur when the system optimizes incorrect or incompletely specified objectives, which can be treated as serious side effects that require entire model redesign. There are growing efforts in the machine learning community to address many forms of biases and to improve the security for safeguarding against adversarial attacks~\cite{kurakin2016adversarial,barocas2017problem,gleave2019adversarial,peng2019you,galhotra2021learning}.

\section{Conclusion}
This article examines the concept of negative side effects of AI systems and offers a comprehensive overview of recent research efforts to address the challenges presented by side effects. In doing so, we aim to advance the general understanding of this nascent but rapidly evolving area. We present a taxonomy of negative side effects, discuss the key challenges in avoiding side effects, and summarize the current literature on this topic. 
This article also presents potential future research directions that are aimed at deepening the understanding of the problem. While some of these issues can be addressed using problem-specific or ad-hoc solutions, developing general techniques to identify and mitigate negative side effects will facilitate the design and deployment of more robust and trustworthy AI systems.

\section*{Acknowledgments}
This work was supported in part by the Semiconductor Research Corporation under grant \#2906.001.

\bibliographystyle{aaai}
\bibliography{References}

\end{document}